*Original Article*

# Maneuvering Digital Watermarking In Face Recognition

Osama R.Shahin[1], Zeinab M. Abdel Azim[2], Ahmed I Taloba[3]

[1,3]Department of Computer Science, College of Science and Arts at Quryyat, Jouf University, Saudi Arabia.
[2]Department of management information systems, Egyptian institute of Alexandria Academy for Administration & Accounting, Alexandria, Egypt.

[1]corresponding.Orshahin@ju.edu.sa, [2]Zena_fmaz@yahoo.com, [3]aitaloba@ju.edu.sa

***Abstract -*** *The challenges faced in the digital world are many, which could be resolved with some biometric recognition methods. These biometric recognition methods are encompassed within watermarking technology, steganography, cryptography, and many other schemes of security. These methods assist in securing digital images with the authentication of their owner. This paper briefly contextualizes the digital watermarking technique, which is referred to as Natural Preserving Transform (NPT) and Hartley Transform, which is endeavored in the face recognition process. The non-blind extraction and quasi-blind extraction techniques are used for extracting the watermark from the image in the proposed system. This paper articulates the application of this watermarking technique employed in face recognition through watermarking of various face images to multiple backgrounds encompassing gray-scale images. Natural Preserve Transform is employed as a part of the fuzzy logic watermarking. In the proposed system, NPT is employed for encoding a logo of grayscale watermarking text or logo image to a host image located anywhere. The robustness and performance of the proposed system are experimentally tested with the help of image processing operations like image compression, noise degradation, cropping. Due to its unique feature of uniform distribution of face images, this technique is selected among other methods in digital watermarking. The system is tested for its efficiency with experimental analyses, which could be confirmed with the results of the simulation. The above system is proposed for copyright protection, authentication, and security requirements.*

***Keywords -*** *Digital watermarking, face recognition, Natural Preserve Transform (NPT), Hartley transform.*

## I. INTRODUCTION

person's physical characteristics are together referred to as his identity, such as a password and signature [1]. The measurement of these physical characteristics is done utilizing various biometric aspects like fingerprint or voice, face, iris, etc., which could be utilized successfully in multiple security-based endeavors like access control, authentication, and identification [2]. The mechanisms related to security have been developed as user-friendly aspects in many endeavors with the aid of biometrics. These mechanisms protect us from identifiers and cryptographic keys [2]. In the proposed paper, digital image watermarking performed with Natural Preserving Transform and Hartley transform is discussed. A procedure where the information is entered as a watermark within an image and which could be detected or extracted later for identification and authentication endeavors is referred to as digital watermarking. This technique is developing as an eminent application because of its feature of accumulation of images in various websites and also because of the imaging industry. This paper encapsulates a grey-scale watermarking technique, where a logo or a watermark is embedded into a host image. The proposed system is a visually meaningful and convenient one with a numerical sequence at random. A watermark encompasses three eminent characteristics such as capacity, visibility, and robustness. A trade-off is achieved with the watermark, which is embedded in the host image [4]. The proposed technique is contemplated on fusion principles of multi-resolution for embedding the gray-scale logo image to the host image in unique blocks encompassing sub-band decomposed wavelets of the host image. The efficiency of the system is tested against cropping attacks. Earlier, the watermarking technique was contemplated on curvelets and was proposed during the embedding of the grey-scale logo. The results of its implementation were not impressive. It is because the normalized correlation (NCORR) between the extracted and original logos was not greater than 0.91. Here a novel watermarking technique is proposed. This technique is contemplated on the utilization of Natural Preserve Transform (NPT). This technique is an orthogonal transform, which is utilized for coding and reconstructing the hidden portion of the signals. Binary logos were utilized by the earlier watermarking schemes. The grey-scale logo in the proposed system is watermarked with even distribution in the host image by the Natural preserve Transform (NPT) evenly around the host image. During extraction of the watermark, this technique requires the host image's prior knowledge. On the extraction side, a quasi-blind and non-blind watermarking technique is proposed.

## II. Objectives and Motivation of the Work

digital images hold thousands of secret and important data, be it personal or medical images. Accordingly, many works have been carried out to process the images and discover the information they contain, whether medical images [5-11], face recognition [12, 13], or vehicles [14, 15]. Therefore, these images need a way to protect them from the information they contain.

Digital watermarking could be encapsulated within a host image through frequency or facial domain. The spatial domain contemplates the visual system of humans for propelling some alterations in the host image, which is not observed [16]. This could be attained with the variation of values in the pixel values' significant bits that are low. In the process of digital watermarking, the watermark is observed with the observation of the addition of the host image and the watermarked image, and the watermark could be extracted enduringly. Generally, the association between visibility and robustness of a watermark in the spatial domain is a vague approach [17]. The maximum host image regions have to be resolved by the user, where a visible watermark is generated as a protective factor from attacks. Data hiding and watermarking are also included in the frequency domain with the help of wavelet transforms or DCT techniques. The initial process in the watermarking technique is embedding, where the transform coefficients for hosting the watermark are chosen for creating invisible watermarks. High-frequency coefficients are less protective when compared with the low-frequency coefficients that exhibit high robustness. The watermarking of high-frequency coefficients is less visible when compared with the low-frequency coefficients. These factors of discrepancy must be considered during the fabrication of a watermark, which is invisible. Robustness has lacked in the watermarking performed on the frequency domain during geometric assails like linear geometric transformation, cropping, and Rotation, Scaling, and Translation (RST). A Natural Preserving Transform (NPT) technique with Hartley transforms is endeavored in the proposed system, encompassing digital image watermarking. It is an iterative extraction procedure, where it encompasses 150-200 iterations for extracting the digital watermark. It is also a slow process. This paper maps out the embedding and extraction procedures of digital watermarking utilizing the Natural Preserve Transform (NPT) and Hartley Transform technique. Here the image with the watermark is constructed with this method, which encapsulates the host image's original part, which is replaced with the logo. This completely hides the watermark. It is a rigid algorithm, which is contemplated on the linear equations having solutions of least squares and which helps in the retrieval of the watermark in the digital image. Results of the paper indicate that the proposed method is a simple, robust, and well-performed system.

## III. Related Work

Face recognition indicates a recognition method, which utilized for detecting the individuals' faces, and these images are saved as a set of data. Face recognition is a significant research topic as it is non-meddling in nature and due to its facile technique of identifying people for their personal identification. In addition, face recognition is one of the successful endeavors in image processing. It is because of the presence of feasible technologies encompassing mobile solutions. Various researches were conducted in automatic recognition of faces based on skin detection, geometrical approaches, eigenvalues that may be used on the recognition of other things such as vehicles [8]. Significant advancements in face recognition have been observed in the past ten years. In spite of these challenges, still, there is some complexities reliability in face tracking and detection and in the recognition of facial patterns. The face recognition method in three dimensions utilizes 3D sensors for capturing information in the face shape. Then this information is utilized for identifying the unique features on the face surface like chin, nose and eye sockets, etc. face recognition is considered an eminent topic for research as it encompasses computer entertainment, computer management modeling techniques, face analysis, verification of face identity and increase in security concerns of the public. Some of the face recognition techniques are the template-based model, appearance-based model, holistic model, geometric-based model.

Face recognition system identifies or verifies individuals utilizing their faces. People in real-time images, videos, photos, etc., are identified with face recognition systems. Mobile devices are used by law enforcement for identifying people with the assistance of the police. The data relating to face recognition may encompass errors, which could affect the people who are crime-free [19]. Sometimes the software related to face recognition may indicate an error while recognizing young people, women, ethical minorities, Americans, and Africans and often failing or misidentifying for identifying them by affecting specific groups. Face recognition systems were utilized for targeting people who are associated with protected speech. Face recognition technology in the future will be ubiquitous. The movements of individuals could be tracked utilizing face recognition technology. Also, this technology is utilized in tracking vehicles by their plate numbers and license numbers. Face recognition in real-time is utilized by sporting events and by different countries.

Computer algorithms are utilized by face recognition systems for picking out distinctive and specific details about the face of a person. These encompass details about the chin or eye shape, eye distance, etc. These details are then changed to a mathematical form and compared with the facial information stored in the database of facial recognition. Face template encompasses details about the facial data about a particular face. A photograph is different from this as it encompasses details about only the differentiation between one face and another face. The probability match score is calculated by some facial recognition methods and not optimistically verifying an unknown person's details. These facial recognition systems provide multiple results like correctness of identification, the rank of identification, etc. The difference in facial recognition systems was identified with their ability to verify people under complex conditions like because of sub-optimal angle, image resolution of low quality, poor lighting conditions, etc. In case of errors, two concepts have to be understood. The mismatch of faces with comparison in the database indicates "false negative", where zero return results are got during queries. "False positive" indicates errors in face recognition systems, which are actually incorrect.

Digital watermarking refers to a technology where the information related to identification is inserted into the carrier of data in unnoticeable ways and where the usage of data would not be impacted. Multimedia copyright data are protected with this technology; also, the text files and databases are protected with the help of the digital watermarking technique. Two basic components are encompassed in a digital watermarking system, such as; the decoder and the encoder [20, 21]. For combining a digital document with a watermark, for instance, in images watermarking, the requirements are an image, which is denoted as Co, a watermarking data contained watermark, which is indicated as 'W', an encoding algorithm, E for producing a watermarked image, indicated as CW and a security key, K. The signature and the cover document is taken by the encoder, and a watermarked image is created. This function is defined as follows:

$$C_W = E(K, W, C_o) \tag{1}$$

Here the public or secret keys and related other parameters are utilized for the extension of the encoder involved in the watermarking process. The watermark would be a robust watermark, and it is inserted in a way where the data CW could survive very serious distortions. A two-step process is involved in a watermark detector or extractor. Retrieval of the watermark is the initial step, which is endeavored for extracting a sequence of data, and they are called retrieval watermarks, which are got from scrambling procedures. The next step is the detection of the embedded watermarks and extraction of the watermark from the signal, which is suspected. This step needs comparison with the original standards, and errors would be detected in this case. The depiction of the watermark detection process is indicated as:

$$W'' = D(K, C_W...) \tag{2}$$

Here, D = Detection method, K= Security Key, W = Hidden signature and $C_W$ = Image Corrupted.

The first process is facial detection, where the camera captures and situates the facial image within the crowd or alone. The image reflects the profile or straight looking of a person. The next step is facial analysis, where the face image is analyzed and captured. Many of the facial recognition technology contemplates on 2D images than the 3D images as the 2D images could be conveniently matched with the pictures of the face recognition database. The face geometry is read by the face recognition software. The features include details about the chin, ears, lip contours, cheek bones' shape distance between chin to forehead, depth of the eye sockets, and distance between two eyes. The next step is the face capture process, where the face or the analog information is converted into the data or the digital information contemplated on the facial features of a person. Then the facial analysis is converted into mathematical form. This mathematical form is referred to as a face print. Similarly, there are unique thumbprints, where face prints are encompassed within each person. Then a match is found for the faces. The face print of a person is then compared with the information in the face recognition database. For example, when the name of a person is tagged in the photo of the person on Facebook, then it becomes a portion of digital data in the database of Facebook.

Digital image watermarking indicates an embedding process, where information is encompassed in digital signal and is being utilized for verifying and identifying the authenticity of its users as similar to the watermark used on paper for identifying visual objects. In the proposed system, the face image is the signal. Copying the signal also copies the information. There is a relationship between digital watermarking and stenography. Digital watermarking also acts as a way for hiding personal information or secrets for protecting the copyright of a product or data, demonstrating integrity. In the digital image watermarking, which is invisible, information is hidden in the visible text or logo or face, which is the face in the proposed system. This information helps in verifying the owner of the message, which is watermarked within the facial image [22].

Natural Preserving Transform (NPT) is one of the techniques involved in digital watermarking. In this technique, the text or logo images involved in the watermarking are grey-scale text or image, where they are watermarked into the host image in the chosen location. NPT fetches a striking feature, where the grey-scale text or logo image is distributed uniformly within the host image in the background in a manner, which could not be perceived. The initial utilization of the Natural Preserve Transform (NPT) was as an orthogonal transform, which encompasses some unusual characteristics, which could be utilized for reconstructing and encoding hidden data within the images. The watermark is extracted with either quasi-blind or non-blind technique in Natural Preserve Transform (NPT) [23]. Only a little information is essential in the quasi-blind technique, as it already has the information in the watermark. The host image is extracted finally with huge data. NPT is an iterative technique involving 150-200 iterations in the watermark extraction. The Natural Preserve Transform (NPT) for an image S of size n*n is given by the following equation:

$$S_t = S_\psi(\alpha) * \psi(\alpha) \tag{3}$$

It is a relatively slow process having rigid least squares. The definition of Natural Preserve Transform could be indicated mathematically as follows:

$$\Psi = H(1-\alpha) + \alpha I_N \tag{4}$$

Where $I_N$ is the identity matrix of the $N^{th}$ order, and H indicates the Orthogonal Hartley Transform.

The Hartley Transform H in two dimensions is indicated as follows:

$$H(k, j) = [\sin(2\pi(j-1)/N) + \cos(2\pi(k-1)/N)] * 1/\sqrt{N} \tag{5}$$

An alternative transform, which encompasses a real-value to the Fourier Transform, is the Hartley transform. It is an efficient tool employed in data analysis. One among the benefits of a Discrete Hartley Transform (DHT) is its similarity with the inverse and forward transforms and is independent against the normalization constant. The inverse and forward Hartley Transform could be implemented by hardware or subroutine with the proper care of the normalization constant. The two-dimensional image data could be transformed utilizing Fast Hartley Transform (FHT). Since real values are carried by Hartley Transform, there is no requirement for complex functions [26-30]. The following shows the algorithm or steps employed in employing Hartley Transform in face recognition systems.

- The image has to be resized initially to an average size of 128*128 and must be transformed to a grayscale image.
- Then the Hartley Transform must be applied to the grayscale image.
- Fractional coefficients must be chosen with the choice of four squares of equal sizes from the transformed image's every corner. These squares must be of the same size.
- The image's feature vector is got by employing Hartley Transform. One whole image is got by the combination of four equal squares of the same size.
- Similar sour steps are employed for the test image as well.
- Then the Euclidean distance between the test image and the feature vectors is calculated, along with the trainee images. The image with the lowest Euclidean distance is chosen as the analogous image. These steps are repeated with the difference in the square sizes of the selected image. Hartley transform is employed in the face recognition database having multiple images. This database is classified into two sets such as the test set and the trainee set.

The Hartley transform is indicated mathematically with the function f(x) as follows:

$$H(s) = \int_{-\infty}^{\infty} cas\, 2\pi xs.\, dx.\, f(x) \tag{6}$$

Cas is presented by Hartley in the year 1942 for standing in place of sine and cosine functions and is given by:

$$cas\, x = sin\, x + cos\, x \tag{7}$$

Hartley transform is an inverse real-valued transform, which indicates the real function as follows:

$$f(x) = \int_{-\infty}^{\infty} cas\, 2\pi xs.\, ds.\, H(s) \tag{8}$$

The eminent step in a face recognition system is the extraction of features of the face image. In the face recognition system contemplated on transform, partial and full feature vectors are taken into consideration. Hartley transform could be applied on images having an average size of 128*128. The partial vector feature is chosen with the cropping of equal square sizes from the corners on four sides of the transformed image. The accuracy got from Hartley Transform is compared with the accuracy got from the Natural Preserve Transform (NPT). The same accuracy and complexity in the computation are given by Natural Preserve Transform (NPT), and Hartley transform. The discrete Hartley Transform is similar to the Discrete Fourier Transform (DFT) and is independent of the two characteristics such as; the similarity of the Hartley Transform with the direct transform and the real nature of the Hartley Transform, which could be managed separately or complex arithmetic for imaginary and real parts. Hartley transform encompasses real values when compared with complex values.

The experimental result of the application of Hartley Transform in the face recognition system indicates that it is a suitable method to be employed for 2D face images. It has high accuracy with the lowest size of the feature vector. Hartley Transform encompasses 2N2 (N-1) number of additions and 2N3 number of multiplications. The striking difference in the Hartley Transform is the feature vector area. Thus, Hartley Transform is suitable for employment in face recognition systems.

The transform T is embedded to image S of a suitable symmetry, and the transformed image is $S_\psi$, which is associated with the original image as follows:

$$S_o = \psi S_\psi \tag{9}$$

When the value of $\alpha = 1$, the original image indicates the transformed image, and during $\alpha = 0$; then it indicates an orthogonal projection. The peak signal-to-noise ratio (PSNR) for the transformed image is $\log_{10}(\alpha/1-\alpha) * 20$. The transformed image retrieves the original image with the expression:

$$S = \psi^{-1} S_o \psi^{-1} \tag{10}$$

Where $\psi^{-1}$ is calculated as follows:

$$\psi^{-1} = 1/\alpha \left[1 - H\left(\frac{1-\alpha}{\alpha}\right) + H^2\left(\frac{1-\alpha}{\alpha}\right)^2 + \ldots\right] \tag{11}$$

The accuracy could be achieved in the desired level at, $\left|H.\frac{1-\alpha}{\alpha}\right| < 1$. Thus, NPT is a suitable method for data hiding and watermarking.

## IV. PROPOSED SYSTEM

The face image and the host image are embedded into the image in the background by the uniform distribution of the images utilizing Natural Preserve Transform (NPT) and also Hartley Transform. The watermarked image is obtained with the information or watermark embedded into the image in the background, and finally, in the extraction process, the host image and the face image with the watermark are extracted. The face image, which is extracted, contemplates the technique utilized, which are the quasi-blind or the non-blind technique. In the quasi-blind method, a little Peak Signal to Noise Ratio (PSNR) is attained, and this achievement is because of the pre-knowledge lacking at the side of the receiver, which is the eminent objective lacking to be made it possible. The entire operation of the system is explained in this section, which encompasses details about the embedding process of the watermark into the background image, the spreading process, extraction process, and the quasi-blind and non-blind techniques for implementing and analyzing the entire system for attaining desired results.

A grey-scale host image in the background and a face image in grey-scale is utilized here. The face image must be embedded into the host image in the background, thus making a watermark. Then the face image would be extracted from the host image that is watermarked with the help of the quasi-blind method. In the quasi-blind method, the original host image is hidden, and with the help of the non-blind method, the original host image in the background is known. Thus, the quasi-blind method is utilized for the watermarking process, and the non-blind technique is employed in the extraction of the host image.

Fig. 1 shows the block diagram of the proposed system, which encompasses a face image, host image, embedding process, watermark process, the extraction process, and the extracted host image and face image as the output.

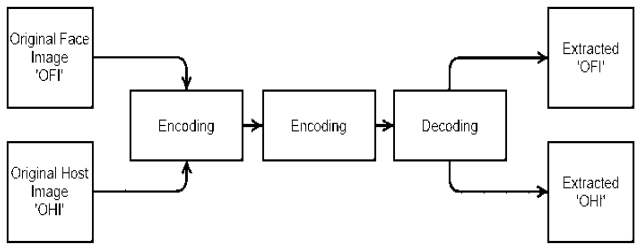

**Fig. 1: Block diagram of the proposed system**

## V. EXPERIMENTAL RESULTS

This section explains clearly the embedding process, the watermarked process, and the extraction process with the quasi-blind method and non-blind method.

### A. *Embedding process*

In the embedding process of digital watermarking, the host image (S) having size N*N has to be watermarked with an m*n size logo image. The embedding of the logo to the host images is done in the embedding technique. The logo to be embedded is a binary watermark in the facial region. Then the reshaping of the logo matrix is done to a matrix p1 having size N*r where r = mn/N. The last rows of the host image are denoted by r, and these rows are replaced with p1. A watermarked square image Spm is got from this step, and it is given by:

$S_{pm}= [s1 p1]$, where S1 = S (1: N-r, :)  (12)

The Natural Preserve Transform (NPT) of S pm is gotten as:

$A_p = \psi(\alpha) \, S_{pm} \, \psi(\alpha)$  (13)

This step uniformly distributes the watermark within the host image in the background. For making the watermarking logo invisible, the last rows, r, are replaced with A p, along with the original image's last row.

$A_{pm} = [A op N : S(N - r + 1), :]$  (14)

This embedding process opts for the replacement of the host image in the background with the logo image. The embedding of the logo is done on the top left corner. The partition of the host image is done as,

$S = [s11 \; s12 s21 \; s22]$  (15)

While the embedding image is done as,

$S_{pm} = [p \; s12 \; s21 \; s22]$  (16)

Then the embedding process continues as below;
- Calculate the NPT of S pm as $A_p = \psi(\alpha) \, S_{pm} \, \psi(\alpha)$
- Partition of the host image is done as follows:

$A_p = [A11 \; A12 A21 \; A22] \; m, N - m$  (17)

- For hiding the watermarking logo, the image which is to be watermarked is A pm, is fabricated with the replacement of the cropping section of the upper left with the original host image, which is given by:

$A_{pm} = [S11 \; A12 A21 \; A22]$  (18)

- Finally, the peak-to-signal ratio is computed between the watermarked image and the original host image in the background for verifying imperceptibility. Then this process is checked for robustness.

In the proposed system, the logo of Zara, a fashion brand, is going to be embedded into the host image. This embedding could be done in any place of the host image. S k is the image region where the logo is to be embedded. This is equivalent to the logo encompassing $\|Sk - p\|$ Euclidean distance. Here we assume embedding the Zara logo into a woman image as shown in Fig.2, utilizing the embedding technique NPT. The size of the fashion brand logo is (n*n), which for instance, is 86*60. The total Hartley orthogonal Transform utilized here is α = 0.991. In the case of bottom embedding, the embedding of the logo is done as; bottom rows =21. Fig. 3 and Fig. 4 indicate the watermarked image with the NPT technique and watermarked image. Apm for the optimum, top, and bottom embedding cases. These three embedding techniques had different values of PSNR, which is given by 39.8dB, 39.6dB, and 39.72dB, respectively. The best watermarking is got with top embedding, and it suits the extraction process as well. This is because of the high energy concentration employed in Hartley Transform. With the embedding of the watermark, the evaluation of the facial image is made. The watermarking applied in the host image is indicated in the below pictures.

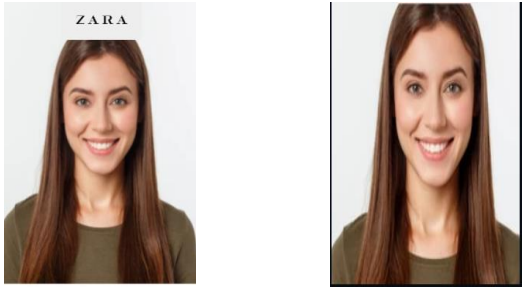

**Fig.2: Optimum embedding with PSNR = 39.8dB**

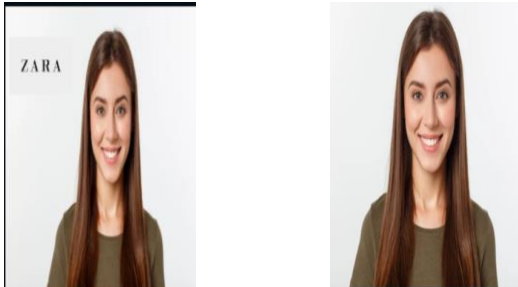

**Fig. 3: Top embedding with PSNR =39.6 dB**

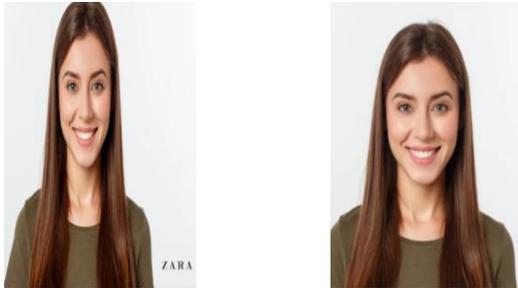

**Fig. 4: Bottom embedding with PSNR = 39.72dB**

*B. Extraction process*

When the host image, s is not present, the quasi-blind method is employed for extracting the watermarked image. This technique involves the blind extraction of the logos, which is embedded in the bottom. The description of the proposed system is indicated as follows.

● Partition image, $\psi = [\psi_{11} \ \psi_{12} \ \psi_{21} \ \psi_{22}]$ (N-r), r
This gives $A_{op}$, $\phi = S_1\psi_{11} + p_1\psi_{12}$

● An (N-r) square matrix is constructed by expressing vector $V_k$ of the $k^{th}$ vector, which is indicated as follows:

$$V_k = I_{k, N-r} - \sum_{j=1}^{r} \psi_{12}(:,j).\alpha_{jk} \quad (19)$$

$$I_{k, N-r} = I_{N-r}(k, i), \text{ where } N-r \geq k \geq 1 \quad (20)$$

In the quasi-blind case, the extraction quality is analyzed with the computation of normal correlation (NCORR) between the extracted logo image and the original image, which is indicated as follows,

$$N_{CORR} = \sum_{i=1}^{m} \sum_{j=1}^{n} a_{exij}a_{ij}/\|a\|.\|aex\| \quad (21)$$

Here, an $_{ex}$ is the extracted watermark. For this process, the host image is an ordinary camera image, and the logo image is a resized image. The size of the camera image is 256*256, and that of the resized image is 32*32. Embedding and reshaping are done on the last four rows of the camera image. The results of extraction showed a watermarked image having $N_{CORR}$ as 1 and PSNR as 32.59db, and the α value is 0.991. Finally, the watermarked image is tested for its robustness against noise, compression, and cropping. The entire Hartley orthogonal transform technique is utilized with α = 0.991.

Fig. 5 indicates the extraction of the watermark, which was embedded at the top of the host image.

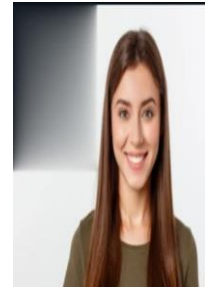

**Fig.5: Extraction at the top embedding**

The extraction of the watermark is classified as the non-blind case and the quasi-blind case. In the non-blind case, the receiver side has information about the original host image, and the trial is made only for extracting the watermarked logo from the image. In the quasi-blind case, only less information about the host image is known on the side of the receiver, and a trial is made for extracting the logo and host images from the image having the watermark, which is $A_{pm}$. The watermark, which is extracted, is computed by the subtraction of the predicted pixel and the watermarked pixel. The results indicated high equality between the watermarked logo and the watermarked image against the compression and cropping attacks.

### VI. Non-blind Extraction Case

In the non-blind case, the trial is made initially for extracting the logo from the image, which is watermarked and embedded at the top. The extraction during the non-blind case requires the knowledge of the parameters such as α, original image S t, Hartley orthogonal transform H N, and watermark extraction parameter A pm. This is further carried

out as below:
- The logo size m*n is determined. This is performed by the correlation of the watermarked image A $_{pm}$ with the host image S for measuring the exact matching region.
- The form for extraction is indicated as follows;

$Y = A_{pm} \phi = [Y11 Y12 Y21 Y22]$ (22)
$Y = S_{pm} \psi$ (23)
$Y = \psi [pS12 S21 S22]$ (24)

Because of the replacement of S $_{11}$ in the location of A $_{11}$, the element Y $_{11}$, Y $_{12}$ sub-matrices, which are non-watermarked, while Y $_{22}$ and Y $_{21}$ would indicate the effects of the watermark, Fig. 6 indicates the logo of Zara, which is to be indicated in the host image.

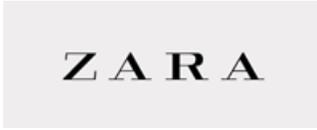

**Fig.6: Logo to be inserted in the face image**

- The partition is indicated as follows:

$\Psi = [\psi11 \psi12 \psi21 \psi22]$ m, N-m (25)

When N-m ≥ m, watermark p is the system's least-square solution.

p. $\Psi_{21} = Y_{21} - S_{21}. \Psi_{22}$ (26)

Though the corruption of Y $_{22}$ did not occur, the calculation of the logo p is not essential in the non-blind case. The extraction technique is linear and efficient computationally in three cases, while it is convergent, fast, and accurate. When there is an equal or less numb, the watermarked logo should be restricted in size, which means that the rows' number must be greater than N/2 for the optimum bottom and top embedding cases.

### VII. The quasi-blind extraction case

When only less information is available about the host image, the quasi-blind method is employed for the NPT contemplated watermark image extraction. For ease, assume embedding of logos in the bottom during the quasi-blind extraction. The description of the proposed method is indicated as follows:

$\Psi = [\psi11 \psi12 \psi21 \psi22]$ N-r, r (27)

As $S_{pm} \Psi = A_p \phi$, it is indicated that,

$[A op S (:, N: N - r + 1)] \phi = [S1 p1][\psi11 \psi12 \psi21 \psi22]$
$A_{op} \Psi = S_1 \Psi_{11} + p_1 \Psi_{12}$ (28)

For canceling the S $_1$ effect in the above equation, (N-r) square matrix is constructed, where L'$\Psi_{12} = 0$. The construction of the square matrix could be expressed in the k$^{th}$ vector as below;

$L_k = I_{k, N-r} - \sum_{j=1}^{r} (:,j) \Psi_{12} \alpha_{jk}$ (29)

Here, $I_{k, N-r} = I_{N-r} (:, k)$. Where N-r ≥ k ≥ 1. R linear equations are solved to get $\alpha_{jk}$ by fulfilling the condition, $L_k^t \Psi_{12} (:,j) = 0$, where r ≥ j ≥ 1. The rank of $\Psi_{12}$ is r, where N-r = $\Psi$. Similarly, the rank of matrix L = N-2r. By pre-multiplying the portioned equation for getting L$^t$ for yielding,
$L^t A_{op} \Psi = L^t S_1 \Psi_{11}$ (30)

A unique solution with r arbitrary values is got in every element of S $_1$, which is known at the side of the extractor or the receiver. The logo p$_1$ is extracted by getting a unique solution of S $_1$ in the non-blind extraction case, and it is slowly reshaped for regaining the original watermark p. For the optimum and top embedding procedure in the quasi blind case, firstly, the original image, p, is extracted in an equal manner, and finally, the logo p $_1$ in the non-blind case is extracted. The knowledge of each column in the r parameter has to be known in the side of the extractor or the receiver for both optimum and top embedding mechanisms, where the logo image should have m number of rows. This indicates that the logo image's area should be hidden within the host image, which leads to noticeable degradation. Thus, for extraction in the quasi-blind case, bottom embedding is highly recommended. The original image's rows are embedded with their corresponding columns. The watermarked image quality is high when there is no reshuffling of data. Furthermore, better extraction is got with an even energy distribution.

### A. Face recognition embedded with digital Image watermarking

Digital image watermarking provides protection and security in the face recognition system. These systems could be used together for maximizing security and recognition performance in an unknown area. When the capturing of the face image is done, it is analyzed for the watermark. When there is a presence of watermarks in this image, then it indicates the presence of some attacks like replay attacks. When there are no watermarks in the image, then the face selection technique is employed for selecting small specific spatial partitions for watermark embedding. This system has endeavored in forensic and biometric applications such as security, performance, and computation time, and resources identification. The discriminative information of the face image is examined in the face images' facial features. Finally, the face system is verified and identified. The face image, which is watermarked, is finally given to the receiver with the integration of all watermarked and non-watermarked partitions within the image. This could enhance the face recognition system and watermark at the side of the receiver. During the investigation of the discriminative information about the face, initially, the high-frequency sub-bands that are discriminative were examined in the face recognition system. Then the study of features of face images such as lip, nose, left eye, and the right eye was examined in the face recognition system. Finally, the investigation of the constructed selection part was carried out with the analysis of its recognition performance for the face identification and face verification systems. The digital image watermarking combined with a face recognition system in an unknown channel is depicted in Fig. 7. Here, the watermark is checked suitable for the face image, and then a face selection method is chosen. The image is watermarked with the face selection method, which is chosen. The watermarked image is communicated to the receiver with a suitable communication channel. The watermark is embedded and extracted, and finally identified or verified with the output got from the receiver. Separate evaluation of the background part and face part is essential during the performance evaluation of the face recognition system.

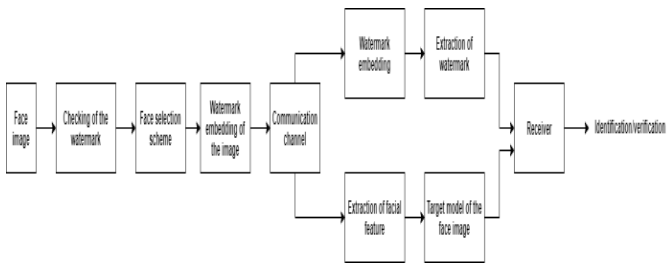

**Fig. 7: Digital image watermarking combined with face recognition system in an unknown channel**

The face size effects were studied with face recognition performance. Various face tests could be employed in examining facial features. One face test is carried out to learn a face image's whole part. The key feature in digital image watermarking is the facial feature than the number of facial features. The face selection methods could choose partitions with low facial features for lowering the effect of degradation of the scheme of image watermarking, cost, and computation time. On the perception of memory overhead and computation time, various schemes of image watermarking have been employed in the extraction and embedding methods of watermarking.

## VIII. CONCLUSION

Thus, this paper concludes about the embedding and extraction of watermarks in face images utilizing the Natural Preserve Transform (NPT) technique encompassing non-blind and quasi-blind methods for extraction of the watermarked image. It is an invisible watermark with robustness against noise, compression, and cropping attacks. A rigid and effective embedding and extraction algorithm are suggested for quasi-blind and non-blind cases of the proposed system. The knowledge about the host image is essential in the non-blind case, and only a little information about the host image is essential in the quasi-blind case. The proposed system was tested, and the results produced an extracted watermarked image, which is perfect and reliable. Even though many security problems could be solved by digital watermarking employed in face recognition systems, some specific facial features are affected. The face recognition system's performance could be improved by taking into consideration the ability of discrimination of facial features lacking uniformity of existence within the image. The watermark availability within the face recognition system is studied during the same time. There was an improvement in time and memory, and improvement is lacking in the recognition rate. Furthermore, facial features and watermark coexistence are confirmed. The experimental evaluation of the proposed system indicated a high efficiency in the application of digital image watermarking in face recognition. Thus, computation time, memory, and performance could be enhanced by endeavoring digital image watermarking in face recognition systems. Thus, the proposed system achieved an effective technique of face recognition with digital image watermarking. Digital watermarking could also be employed in theft protection of original facial images and in marketing endeavors. Tracking components are encapsulated within the images, which have information about the original image. Thus, data and databases are secured utilizing the digital watermarking technique. Further research is essential for the proposed system in security applications and law enforcement.